\documentclass[twoside, 12pt]{article}
\usepackage{pgfplots,comment}
\usepackage[a4paper,top=1.18in, bottom=1.18in, left=1.2in, right=1.2in,bindingoffset=0mm]{geometry}

\usepackage{aditya2,setspace}

\tikzset{
    >=latex,
    pil/.style={
            draw,
      <-, 
      decorate,
      decoration={snake,,amplitude=.02cm, pre length=.2cm,post length=.2cm,}
              }}

\hypersetup{colorlinks=true,linkcolor=red, citecolor= blue, urlcolor=darkred}

\definecolor{BluishGreen}{RGB}{0,158,115}

\title{{\bf \sc {{\color{darkblue}matching with incomplete preferences}}}}

\author{
\begin{minipage}{0.3\textwidth}\centering 
Aditya Kuvalekar\footnote{Email: \texttt{a.kuvalekar@essex.ac.uk} \\I am indebted to Kazuhiro Hara and Elliot Lipnowski for their thoughts and help on this paper. I would like to thank Nageeb Ali, Joyee Deb, Laura Doval,  Bhaskar Dutta, Piotr Dworczak, Mikhail Freer, Niccol\`o Lomys, Ana Mauleon, Marek Pycia, Ariel Rubinstein, Arunava Sen and Alex Teytelboym for comments and suggestions.  }  \\ \centering \it \small University of Essex
\end{minipage}                  
}

\date{\vspace{0.0cm} October, 2023}

\begin{document}
\vspace{-1cm}
\maketitle

\vspace{-0.4cm}
\begin{spacing}{1.15}
\begin{abstract}

{\small \noindent} 
I study a two-sided marriage market in which agents have incomplete preferences---i.e., they find some alternatives incomparable. The strong (weak) core consists of matchings wherein no coalition wants to form a new match between themselves, leaving some (all) agents better off without harming anyone. The strong core may be empty, while the weak core can be too large. I propose the concept of the ``compromise core''---a nonempty set that sits between the weak and the strong cores. Similarly, I define the men-(women-) optimal core and illustrate its benefit in an application to India's engineering college admissions system.
\\

\bigskip

\vspace{0.2cm}
\end{abstract}

\section{Introduction}
In market design, it is typically assumed that agents have complete preferences---they can compare any two options. Several authors have questioned the assumption of completeness in various contexts.\footnote{For example, \cite{aumann1962utility}, \cite{bewley1986knightian}, \cite{ok2002utility}. } Especially in applications such as school choice, 
incomplete preferences may arise naturally. 
Consider the Indian Institute of Technology' (IIT) admission system. 
Each year, $1.3$ million students seek admission for about $34,000$ engineering seats across IITs and a few non-IITs. The current procedure involves ranking students based on their performance on two exams (\cite{baswana2019centralized}). 
Thereafter, students submit their preferences---a strict, complete ordering over available choices.
A seat is a two-dimensional object consisting of the institute---e.g.\ IIT Bombay, IIT Delhi---and the discipline of study---e.g. computer science, mathematics, chemical engineering. Suppose  Anita prefers IIT Bombay over IIT Delhi (the former being closer to her home), whereas her preferred discipline is computer science, followed by chemical engineering, followed by mathematics. 
Then, Anita might find computer science at IIT Delhi \emph{incomparable} with chemical engineering at IIT Bombay. Many students face this dilemma each year---making a choice between a more-preferred institute and a more-preferred discipline---and a number of websites offer advice to students on these dimensions.\footnote{For example,  
\url{https://www.quora.com/What-branch-and-IIT-should-I-choose}. }

When forced to report complete preferences, students must artificially resolve this indecisiveness one way or the other. However, might it be ``better'' to allow students to express incompleteness? How should we think of stability, and the core, when preferences are incomplete?  These questions motivate the current paper.

I study the two-sided marriage market problem as in \cite{gale1962college}, but with one distinction---agents have \emph{transitive but possibly incomplete} preferences.\footnote{An important question is whether the distinction between indecisiveness and indifference is important. Why can we not just replace indecisiveness with indifference? I discuss this at the end of the introduction.}
With incompleteness, two natural notions of the core (and stability) immediately come to mind. The weak core consists of matchings wherein no coalition wants to deviate and form a match between themselves that improves the allocation of \emph{all} the agents in the coalition. The strong core consists of matchings wherein no coalition wants to deviate and form a match between themselves that improves the allocation of \emph{at least one} agent in the coalition \emph{without harming anyone} in the coalition.\footnote{Notice that, in contrast to the situation with complete preferences, some agents participating in a coalitional deviation here may move to an allocation that is incomparable with their original allocation. See Definition \ref{Definition: the strong core} for a formal version.} 

When agents have incomplete preferences, the strong core may be empty and the weak core may be too large. 
However, these problems are not the sole motivation for an intermediate solution concept. Consider the example below. 

\bexam\label{Example intro: nontransitivity of strong dominance}
Let $M = \{m_1,m_2\}$ and $W=\{w_1,w_2\}$.
Each $w_i$ likes $m_1$ better than $m_2$. However, women cannot compare remaining single with being matched with either man. 
Men cannot compare any alternatives. 
In particular, they cannot compare staying single with being matched.
\eexam 
The weak core consists of all possible matchings which we enumerate below, where an ordered pair denotes a match. For example, in $\mu_1$ all the agents are single, while in $\mu_2$, $m_1$ and $w_1$ are matched to each other, and $m_2$ and $w_2$ are single. 	
\begin{eqnarray*}
\mu_1 = &\{\}, & \mu_2 = \{(m_1,w_1)\}, 
\mu_3 = \{(m_2,w_2)\}\\
\mu_4 = &\{(m_2,w_1)\}, &\mu_5 = \{(m_1,w_2)\},  \mu_6 = \{(m_1,w_1), (m_2, w_2)\}\\
\mu_7 = &\{(m_2,w_1), (m_1, w_2)\}
\end{eqnarray*}

The strong core is $\{\mu_1,\mu_2,\mu_5\}$.

I claim that the weak core is too permissive, while the strong core is too restrictive in the above example. Notice that the weak core contains even $\mu_3$ and $\mu_4$---matchings wherein each woman is assigned to her second-ranked man, while the top-ranked man remains single. This seems somewhat unsatisfactory. The strong core justifiably eliminates those, but, in doing so, it also eliminates $\mu_6$ and $\mu_7$---matchings wherein all the agents are matched. 

Compare $\mu_6$ with $\mu_2$, for instance. The only difference between the two is that $\mu_6$ matches $m_2$ with $w_2$ while $\mu_2$ keeps them single. More importantly, neither $m_2$ nor $w_2$ can compare remaining single with being matched to each other. 
Therefore, as an analyst, we may not want to necessarily assert that $\mu_6$ is better than $\mu_2$ just because it forms more matches. However, by including $\mu_2$ while excluding $\mu_6$, the strong core implicitly takes a strong stand \emph{against} forming matches in the above example.\footnote{This is not to suggest that the strong core takes a stand against forming matches more generally.}

In short, even when the strong core is nonempty, it may refine the weak core to an unappealing subset. Can one do ``better''? If yes, in what sense?  The main contribution of my paper is to answer these questions by proposing a concept of the core: ``\emph{the compromise core}''---a nonempty set that sits between the weak and the strong cores. 

As the name suggests, it is a compromise that circumvents the emptiness of the strong core  while retaining the power to refine the weak core. Moreover, Theorem \ref{Theorem: Characterization of C} provides the sense in which the compromise core is a meaningful refinement: it is characterized by three normatively natural axioms. 
In Example \ref{Example intro: nontransitivity of strong dominance}, the compromise core selects $\mu_6$ and $\mu_7$ along with the strong core.

The compromise core is predicated on the concept of ``strong moves''---moves by coalitions  where a subset of agents inside a coalition become strictly better off without harming anyone inside the coalition. Strong moves---also responsible for the potential emptiness of the strong core---motivate an order called ``dominance'' on the weak core.
Loosely speaking, a matching $\mu$ dominates $\mu'$  if a coalition can precipitate a move from $\mu'$ to $\mu$ (can redistribute matches among themselves without disturbing the matches outside), and some members of the coalition find such a move strictly desirable, while no member finds it undesirable. 

Unsurprisingly, dominance is sometimes intransitive and sometimes incomplete. If this relation were transitive, we could simply select the maximal elements. Therefore, the question is, how should one refine the weak core endowed with a natural but an intransitive order? Incidentally, an extensive literature in social choice theory constructs reasonable choice rules that circumvent the problem of nonexistence of clear winners due to intransitivities (e.g. majority rule in voting). A particularly appealing choice rule is the uncovered set due to \cite{miller1980new}.\footnote{The notions of the uncovered set due to \cite{miller1980new} and \cite{fishburn1977condorcet} are (not) equivalent when the underlying binary relation is complete (incomplete) and strict. } 
 The Miller covering induced by my dominance relation, henceforth \emph{covering}, says that a matching $\mu$ covers $\mu'$ if, $(i)$ $\mu$ dominates $\mu'$ and $(ii)$ $\mu$ dominates any matching that $\mu'$ dominates.
The compromise core is the nonempty set of uncovered elements.

Let me briefly comment on a natural question raised earlier: Is the distinction between indifference and  indecisiveness important? Indeed, if one views choice data as a primitive and preferences as derived objects, then incomparability is indistinguishable from indifference \emph{if we allow for intransitive preferences}.

Let us go back to Anita's example and say that she is choosing among three colleges, $C_1,C_2$, and $C_3$. She ranks $C_2$ better than $C_3$ because it is closer to her home and offers a discipline that she prefers. However, she cannot compare $C_1$ with either $C_2$ or $C_3$ because $C_1$ is the farthest from her home but offers the discipline she most prefers.  Notice that if we replace all the incomparabilities for Anita with indifference, then her preference would be $C_1 \sim C_2, C_1 \sim C_3$ \emph{and} $C_2 \succ C_3$---the preference would be intransitive.\footnote{\cite{luce1989games} discuss the possibility of intransitivities occurring when subjects compare inherently incomparable alternatives.} 
However, at least in the context of school choice, there is an important distinction between indifference and incompleteness. \cite{roth18sotomayor} suggest:
\begin{quote}
Loosely speaking, the reason is that indifference is in some sense a ``knife edge'' phenomenon; if an agent is indifferent between two alternatives, a small improvement in one of them would presumably cause him or her to prefer it to the other. (p.35)
\end{quote}
Translating this into Anita's environment, if, hypothetically, $C_1$ were to move closer to her home by a tiny distance, Anita should strictly prefer $C_1$ over both $C_2$ and $C_3$ \emph{if she were indifferent}. It is the incomparability between these colleges that would prevent Anita from having such a strict ranking. 

While agnostic about the foundation of incompleteness \emph{vs.} intransitive indifference, I adopt the interpretation of indecisiveness throughout. However, my proposed solution concepts and results remain equally valid when preferences are allowed to be intransitive with a transitive strict part. 

Finally, I briefly discuss the IIT admission system. 
I show that the current process may generate matchings in the weak core that are strictly dominated according to students' preference by another matching in the weak core. Therefore,  allowing students to express incomplete preferences may materially improve outcomes.

\section{Model}\label{Section: model}

\subsection{Preliminaries}

Let $M$ and $W$ denote disjoint finite sets, which I refer to as men and women following the literature. Let $A:= M \cup W$. A generic agent in $M$ [resp. $W]$ will be denoted by $m [w]$. When the distinction is immaterial, I denote an agent by $i$. Let $O(i) = W\cup \{i\}$ if $i \in M$, and $O(i) = M\cup \{i\}$ if $i\in W$, denote the \emph{opposite side} for agent $i$.

Each agent $i \in A$ is endowed with a preference relation, $\succsim_i$ over $O(i) $. $\succsim_i$ is a binary relation that is reflexive, transitive, and  antisymmetric, but possibly incomplete.\footnote{A binary relation $R$ on a set $X$ is relfexive if $x R x$, it is transitive if $x R y, y R z \implies x R z$, and it is antisymmetric if $x R y, y Rx \implies x = y$.} 
That is, $i$ may find two agents, $j, k \in O(i)$ incomparable. I denote this by $j \otimes_i k$.  Moreover, since $\succsim_i$ is antisymmetric, $j \succsim_i k$ means that $i$ strictly prefers $j$ over $k$, denoted $j \succ_i k$, whenever $j \neq k$.\footnote{While I take preferences as the primitive, one could, alternatively start with a choice behaviour that is consistent with incomplete preferences as a primitive to define stability and core as in \cite{che2019weak} for example. \cite{eliaz2006indifference} provide a revealed preference foundation of incomplete preferences.} I say $\succsim'_i$ is \emph{``more complete''} than $\succsim_i$, if $j \succsim_i k \implies j \succsim'_i k$ for all $i \in A$. 
Lastly, $\succsim := (\succsim_i)_{i \in A}$ is called a preference profile. 

Let $S$ be a set endowed with a binary relation $R$, which need not be complete or transitive. Let $P$ be the strict part of $R$, i.e. $x P y$ if $x R y$ but not $y R x$. We say that $x$ is a maximal element of $S$ if no $y \in S$ has $y P x$; $x$ is a maximum element of $S$ if $x P y$ for all other $y \in S$. The set of maximal elements is denoted by $MAX(S, R)$. 

\smallskip

A bijection $\mu:A \to A$ is called a matching if $\mu(i) \in O(i) $ for all $i \in A$, and $\mu(i) = j \implies \mu(j) = i$.  I let $\mc M$ denote the set of all matchings. 
Say a matching $\mu \in \mc M$ is \emph{individually rational} if no $i \in A$ has $i \succ_i \mu(i)$. Below, I define some natural notions of stability.

\bdefn\label{Definition: weak stability}
A matching $\mu$ is ``weakly stable'' if it is individually rational and no $(m,w)\in M\times W$ has $w \succ_m \mu(m)$ and $m \succ_w \mu(w)$. If there exists such a $(m,w)$, then I say that $(m,w)$ is a ``strong blocking pair'' for $\mu$. 
\edefn 

\bdefn\label{Definition: weak core}A matching $\mu$ is in the ``weak core'' if $\nexists$ $S \subset A$ and a matching $\mu'\neq \mu$ such that $\mu'(i) \in S$ for all $i \in S$, and $\mu'(i) \succ_i \mu(i)$ for all $i \in S$. If there is such an $S$, then I say that $S$ strongly blocks $\mu$. 
\edefn  
The reason to call the above as a ``strong blocking pair'' is that we could define a weaker notion---a weak blocking pair---as below. 

\bdefn Given a matching $\mu$, a pair $(m,w)$ is a ``weak blocking pair'' for $\mu$ if one of the following holds.
\begin{enumerate}
\item $(m,w)$ is a strong blocking pair. 
\item $m \succ_w \mu(w)$ and $w\otimes_m \mu(m)$.
\item $w\succ_m \mu(m)$ and $m\otimes_w \mu(w)$. 
\end{enumerate} \edefn

In contrast to the strong blocking pair, the notion of a weak blocking pair is somewhat nonstandard. If $(m,w)$ constitute a weak blocking pair for $\mu$, it means the following: by leaving their respective match in $\mu$ and matching with each other,  $(1)$ neither agent is worse off, and, $(2)$ at least one of them is strictly better off. One may wonder why an agent would participate in a move where (s)he switches between two allocations that (s)he finds incomparable. The reason why such moves maybe reasonable stems from the observation that every incomplete preference relation is an intersection of a set of complete preference relations. Therefore, we can view an agent's incomplete preference as if it arises from using multiple criteria to evaluate the alternatives, as in Anita's example from the Introduction. An agent, say $m$, may find two alternatives incomparable because some of the criteria he uses, do not rank the said two alternatives the same way. In such a case, the reason why $m$ may participate in a move from one alternative to the other is simply because he is not worse off from such a move. In other words, participating in such a move assumes acquiescence rather than hesitancy on the part of the agent in moving between two incomparable allocations. 

\bdefn\label{Definition: strong stability}
A matching $\mu$ is ``strongly stable'' if it is individually rational and has no weak blocking pair. 
\edefn

\bdefn\label{Definition: succ_S order} Let $\mu(S) := \{\mu(i): i \in S\}$. Say that $\mu \succ_S \mu'$ if 
\begin{enumerate}[(i)]
    \item $\mu(S) = \mu'(S)$, and,
    \item $\mu(i) \nprec_i \mu'(i)$ for all $i \in S$, and, $\mu(i) \succ_i \mu'(i)$ for at least one $i \in S$.
\end{enumerate}

\edefn 
Notice that, if the preferences were complete, then $\succ_S$ would coincide with the Pareto ranking (for the agents in $S$), namely, $\mu \succ_S \mu'$ if $\mu(i) \succsim_i \mu'(i)$ for all $i \in S$, and $\mu(i) \succ_i \mu'(i)$ for some $i\in S$. On the other hand, with incomplete but strict preferences, matchings undominated according to the Pareto ranking (as defined on the previous line), would coincide with the weak core. Of course, the definition of $\succ_S$ as in Definition \ref{Definition: succ_S order} is less demanding than the Pareto ranking when the agents have incomplete preferences. Similar to the weak blocking pairs, it captures the idea that some agents in $S$ may be willing to participate in a move from $\mu'$ to $\mu$ simply because they are not worse off from such a move.

\bdefn\label{Definition: the strong core}A matching $\mu$ is in the ``strong core'' if $\nexists$ $S \subset A$ and a matching $\mu'$, such that $\mu' \succ_S \mu$. If there is such an $S$, then I say that $S$ weakly blocks $\mu$. 
\edefn

Proposition \ref{Proposition: equivalence of core and stability} below, whose analogues for complete preferences are well-known in the literature, establishes that the weak (strong) core and the set of weakly (strongly) stable matchings coincide. Therefore, I refer to these two concepts interchangeably henceforth. 
\bprop\label{Proposition: equivalence of core and stability}A matching $\mu$ is weakly [resp. strongly] stable iff it is in the weak [strong] core. \eprop

Let $\W $ [resp. $\cs$] denote the weak [strong] core.

A binary relation $\comp$ is a ``completion of $\succsim$'' if, for each $i$, $\comp_i$ is a complete, reflexive, transitive, and antisymmetric binary relation over $O(i)$, such that $j \succsim_i k \implies j \comp_i k$.

When the underlying preference is complete, the notions of weak and strong stability coincide. Therefore, when agents have complete preferences, a matching is stable if it is weakly stable. It is easy to characterize the weakly stable matchings in terms of completions of $\succsim$. To this end, let $C(\succsim)$ denote the set of completions of $\succsim$. 
\bprop\label{Proposition: weak stable iff stable according to some completion} A matching $\mu$ is weakly stable according to $\succsim$ iff it is stable according to some completion $\comp$. 
\eprop 

The following corollaries are immediate. 
\bcor If $\succsim$ is more complete than $\succsim'$, then $\W(\succsim) \subseteq \W(\succsim')$. \ecor 

\bcor\label{Corollary: weak core nonempty} The weak core is nonempty.\ecor

In contrast to Corollary \ref{Corollary: weak core nonempty}, and as one might expect, the strong core may be empty, as the following example demonstrates. 

\bexam\label{Example: empty the strong core}
Let $M = \{m_1,m_2,m_3\}, W= \{w_1,w_2,w_3\}$. All men have identical preferences: $w_1 \otimes_m w_2, w_1 \otimes_m w_3, w_2 \succ_m w_3$ for all $m\in M$. All women have identical preferences, too: $m \otimes_w \tilde m$, for all $w \in W$ and different $m,\tilde{m} \in M$. Moreover, all the agents strictly prefer being matched over remaining single.  
\eexam

First, it is obvious that $\W= \{\mu \in \mc M: \mu(i) \neq i \forall i \in A\}$. Consider an arbitrary matching $\mu \in \W$. Letting $m=\mu(w_3)$, notice that $(m,w_2)$ is a weak blocking pair as $w_2 \succ_m w_3$ and $m \otimes_{w_2} \mu(w_2)$. 

Analogous to Proposition \ref{Proposition: weak stable iff stable according to some completion}, one might wonder whether the strong core coincides with matchings that are stable with respect to every completion of $\succsim$. This is, however, not true. 

\bprop\label{Proposition: strong core not intersection} If $\mu$ is stable with respect to every completion of $\succsim$, then $\mu$ is in the strong core. However, the strong core can be strictly larger than the set of matchings that are stable with respect to every completion. 
\eprop 
To see the strictly larger part, consider a market with one man and one woman such that no agent is able to compare remaining single to being matched. The strong core contains all possible matchings. However, no matching is stable with respect to every completion.

\section{Compromise Core}\label{Section: compromise core}
While the strong core may be empty as we just saw, Example \ref{Example intro: nontransitivity of strong dominance} demonstrates how it can discard intuitively appealing matchings even when it is non-empty. 
These observations motivate the search for an alternative solution concept like the compromise core. 
At its heart are coalitional moves. Weakly stable matchings are prone to \emph{strong moves} by coalitions---moves where only a subset of agents need to strictly improve while nobody inside the coalition is worse off. 
Unsurprisingly, it is possible to cycle through weakly stable matchings when one allows for strong moves. 
Therefore, I begin the process of refining the weak core by endowing the weak core with the dominance order: a matching $\mu$ dominates $\mu'$ if there is a coalition that can ``\emph{enforce}'' a move from $\mu'$ to $\mu$, and the coalition also finds such a move ``\emph{preferable}''---some members of the coalition find this move strictly better off while nobody in the coalition is worse off. 
Below, I make these ideas precise.  

\bdefn\label{Definition: enforceability mauleon} Given a matching $\mu$, a coalition $S \subset A$ is said to be able to enforce a matching $\mu'$ over $\mu$ if the following conditions hold: (a) $\mu'(i) \notin \{\mu(i), i\} \implies \{i,\mu'(i)\}\subset S$ and (b) $\mu'(i) = i \neq \mu(i) \implies \{i,\mu(i)\} \cap S \neq \emptyset$. \edefn

The above definition is due to \cite{mauleon2011neumann}. As they say, ``This enforceability condition implies both that any new match in $\mu'$ that does not exist in $\mu$ should be between agents in $S$, and that to destroy an existing match in $\mu$, one of the two agents involved in that match should belong to coalition $S$.''

Below, I define dominance---an order that forms the basis of refining the weak core.

\bdefn \label{Definition: strong move}Consider two matchings $\mu, \mu' \in \W$.\footnote{The dominance relation can be defined over $\mc M$ and not just $\W$. However, since the objective of the exercise is to refine $\W$, I directly define it over $\W$.} Say that $\mu$ dominates $\mu'$, denoted by $\mu \unrhd \mu'$, if there is $S \subset A$ such that
\begin{enumerate}
\item $S$ is able to enforce a matching $\mu$ over $\mu'$, and,
\item $\mu \succ_S \mu'$.\footnote{Recall that $\mu \succ_S \mu'$ means that no $i\in S$ has $\mu'(i) \succ_i \mu(i)$, and some $i \in S$ has $\mu(i) \succ_i \mu'(i)$.} 
\end{enumerate} 
$\rhd$ denotes the strict part of $\unrhd$. That is, $\mu \rhd \mu'$ if $\mu \unrhd \mu'$ and $\neg (\mu' \unrhd \mu)$. 
\edefn

Notice that predicating a refinement on the dominance relation entails considering blockings that involve incomparable matches. That is, when $\mu \rhd \mu'$, a coalition that enforces a move from $\mu'$ to $\mu$ will have some agents' allocation in $\mu$ and $\mu'$ being incomparable. If these were indifferences instead of incomparabilities, then such moves are easy to justify---the agents who strictly prefer their allocation in $\mu$ can pay a small ``bribe'' (not formally in the model) to induce the indifferent agents to accede to the coalitional move. However, if one adopts a (reasonable) perspective that moves involving transitions between incomparable allocations should not be allowed, then it seems that the weak core cannot be refined further in light of Proposition \ref{Proposition: weak stable iff stable according to some completion}. Therefore, to the extent that one views refining the weak core as a meaningful exercise, the dominance relation offers a reasonable alternative towards that goal.

Dominance, while intuitively appealing as an order on $\W$, can be incomplete or intransitive. 
Moreover, $\unrhd$ need not be antisymmetric either, i.e., there can exist $\mu, \mu' \in \W$ such that $\mu \unrhd \mu' \unrhd \mu$ (Lemma \ref{Lemma: lhd incompelte, intransitive, antisymmetric} in the Appendix).  Also, one may  wonder if $\unrhd$, when restricted to $\W$, is essentially a form of Pareto dominance. That is, say that $\mu$ strongly Pareto dominates $\mu'$ if $\mu \succ_{M\cup W} \mu'$. 
Of course, if $\mu$ strongly Pareto dominates $\mu'$ then $\mu \unrhd \mu'$. The converse is not true. 

If $\unrhd$ were a nice binary relation (e.g. complete and transitive), the task of refining $\W$ would be straightforward: Choose the $\unrhd-$maximal elements. However, due to intransitivities, a maximal element may not exist. 
This is neither surprising (given how dominance is defined), nor is the problem unique---especially for the practitioners of social choice theory. Such intransitivities abound in problems dealing with voting rules where the majority rule, while being complete, often fails to be transitive. As \cite{dutta1988covering} says, one of the popular pastimes of social choice theorists has been to construct choice rules that handle the difficulty posed by the nonexistence of a clear winner. 

One such choice rule that seems particularly well-suited for our environment is the ``\emph{uncovered set}'', proposed in \cite{fishburn1977condorcet} and \cite{miller1980new}. 
To this end, I introduce the notion of covering below.

\bdefn Given $\mu, \mu' \in \W$, we say that $\mu$ covers $\mu'$, denoted by $\mu \ll \mu'$ if, (i) $\mu \rhd \mu'$ and, (ii) every $\mu'' \in \W$ with $\mu' \rhd \mu''$ has $\mu \rhd \mu''$.\footnote{Recall that the definition of dominance, $\rhd$, is given in Definition \ref{Definition: strong move}, while the definition of $\succ_S$ (in the definition of dominance) is given in Definition~\ref{Definition: succ_S order}.} \edefn

While $\rhd$ may itself be intransitive or incomplete, $\ll$ is an antisymmetric and transitive (but possibly incomplete) binary relation (Lemma \ref{Lemma: covering is transitive} in the Appendix). The compromise core, formally defined below, is the set of $\ll$\textendash maximal elements or, the uncovered set: matchings that are not covered by any other matching.

\bdefn\label{Definition: compromise core} The compromise core $\C$ is the set of all the matchings $\mu \in \W$ that are not covered by any other matching $\mu' \in \W$. Formally, $\C := \{ \mu \in \W : \nexists \mu' \in \W \text{ s.t. } \mu' \ll \mu\}$. 
\edefn

\brem A detailed expository example illustrating the compromise core is available in the Appendix \ref{Section: appendix example of a compromise core}.\erem 

Several notions of covering are extensively studied in social choice theory.\footnote{See \cite{duggan2013uncovered} for a detailed discussion and comparison of some variants. Some prominent ones are \cite{gillies1959solutions}, \cite{fishburn1977condorcet}, \cite{miller1980new}, \cite{bordes1983possibility}. } While they vary in their discriminatory power, I find the covering relation due to \cite{miller1980new}  the most appealing. To see why, let us ask why we would want a certain matching admitted by our solution concept? One reason would be that it is undominated. The other would be that it dominates a subset of matchings that makes it compelling. However, if $\mu$ covers $\mu'$, then $\mu'$ is neither undominated nor does it dominate any matching that $\mu$ does not. Therefore, there is arguably no additional advantage of selecting $\mu'$ over $\mu$. 
Consequently, when equipped with the goal of refining $\W$, if one must include one of $\mu$ and $\mu'$, it should be $\mu$.  Theorem \ref{Theorem: Characterization of C} offers further justification for the compromise core through three simple axioms. 

There is another well-studied notion of covering, the Fisher covering (\cite{fishburn1977condorcet}), that is close to Miller-covering, i.e., the notion of covering used in this paper. We say that \emph{a matching $\mu$ F-covers $\mu'$ if, $\mu \rhd \mu'$ and $\mu'' \rhd \mu \implies \mu'' \rhd \mu'$.} If $\unrhd$ were a strict and a complete order, F-covering and Miller-coverings are equivalent. However, especially with the incompleteness of $\unrhd$, the Fisher uncovered set seems unappealing. To see this, suppose that $\W = \{\mu_1, \mu_2,\mu_3\}$ for some matching environment wherein, $\mu_1 \rhd \mu_2$ and $\mu_2 \rhd \mu_3$ but $\mu_1$ and $\mu_3$ are not $\rhd-$comparable (e.g.\ Example \ref{Example: problem with Fishburn covering} in the Appendix is one such instance). Consider the following two properties that one may consider desirable from a refinement. First, any $\rhd-$undominated matching must be selected. Second, any $\rhd-$dominated matching that does not dominate any other matching, must be excluded. That is, if $\mu \rhd \mu'$ and there is no $\mu''$ such that $\mu' \rhd \mu''$, then $\mu'$ should be excluded.   
Putting these two together, the two possible candidate refined sets are: $\{\mu_1\}$ and $\{\mu_1, \mu_2\}$. 
Notice that, $\mu_2$ Miller-covers $\mu_3$ while $\mu_1$ and $\mu_2$ are Miller-uncovered. On the other hand, $\mu_1$ F-covers $\mu_2$. Therefore, the compromise core is $\{\mu_1, \mu_2\}$ while the 
Fisher uncovered set is $\{\mu_1, \mu_3\}$. It is difficult to justify a refinement that includes $\mu_3$ while excluding $\mu_2$.


I wish to point out two structural differences between $\C$ and the Miller-uncovered set defined on a set endowed with a complete order. 
First, when the underlying relation is complete, any alternative in the Miller uncovered set beats any other alternative in at most two steps. That is, given any $\mu$ in the Miller uncovered set and any $\mu'\in\W$, either $\mu \rhd \mu'$ or $\exists \mu_1 \in \W$ such that $\mu \rhd \mu_1 \rhd \mu'$, \emph{if $\rhd$ were complete.} With incompleteness, this is not true. More importantly, whenever the underyling relation is complete, the Miller uncovered set is contained in the top cycle set defined as follows:
\begin{align*}
TC := MAX( \W, \unrhd^T)
\end{align*}
where $MAX(A, R)$ denotes the maximal elements of set $A$ according to a binary relation $R$, while $R^T$ denotes the transitive closure of $R$.\footnote{Transitive closure of a relation $R$ is defined as $x R^T y$ iff $\exists x_1, x_2,\ldots x_n$ such that $x R x_1 R x_2 \ldots x_n R y$. } However, $\C$ can be strictly larger than the top cycle.

\subsection{Relationship to the strong and the weak core}\label{Section: compromise core and farsighted stability}

We now see that the strong core is a subset of the compromise core. 

\bprop\label{Proposition: strong dominance iff the strong core} If $\mu \in \cs$, then $\nexists \mu' \in \W$ such that $\mu' \rhd \mu$. Therefore, $\cs \subseteq \C$. \eprop 

\bprop\label{Proposition: the strong core subset compromise core subset weak core} The strong core is a subset of the compromise core, while the compromise core is a nonempty subset of the weak core. Moreover, these inclusions can be strict even when the strong core is nonempty. \eprop 
\bprf 
By Proposition \ref{Proposition: strong dominance iff the strong core}, $\cs \subseteq \C$. 
The second inclusion, $\C \subset \W$, is an immediate consequence of the definition of the compromise core.
That, $\C$ is nonempty follows from the transitivity of $\ll$ as mentioned before.
The second part of the proposition is shown in Example \ref{Example intro: nontransitivity of strong dominance}. 
Recall that, in that example, $\cs = \{\mu_1, \mu_2, \mu_5\}$, and therefore they belong to $\C$. Finally, notice that $\mu_2 \rhd \mu_7$ and $\mu_7 \rhd \mu_3$, but $\mu_2$ and $\mu_3$ are not $\unrhd$-comparable. Moreover, there is no other $\mu$ such that $\mu \rhd \mu_7$. Therefore, $\mu_7$ is uncovered. Similarly, $\mu_6$ is uncovered. Therefore, $\C = \{\mu_1, \mu_2,\mu_5,\mu_6,\mu_7\}$. 
\eprf
Notice that the top-cycle in the above example is $\{\mu_1, \mu_2, \mu_5\}$ which is strictly smaller than $\C$.

In Example \ref{Example intro: nontransitivity of strong dominance}, the compromise core contains, beyond the strong core, the matchings in which no agent is single. 
On the other hand, matchings that are not in the compromise core match only one woman to her second-best partner, leaving her best partner unmatched.
This highlights an 	advantage of the compromise core over the weak and the strong cores.
The compromise core creates more pairs than does the strong core, to fulfill some individuals by exploiting the incomparability of others.
In doing so, the compromise core is more selective than the weak core.

\section{Normative Foundations of the Compromise Core}
In this section, I show that a set of normatively appealing axioms characterize the compromise core. 


To be concrete, consider the problem of designing a map $\G :2^\W \to 2^\W$ such that $\emptyset \neq \G(A) \subset A$ for all $A \subset \W$. That is, given a set of matchings in $\W$, $\G(\cdot)$ refines them to produce a subset of those matchings. For example, the compromise core, $\C(A)$---the set of $\ll$-uncovered elements in $A$---is one such map. 

Let us introduce some notation. First, $\mu^\uparrow := \{ \mu' \in \W : \mu' \rhd \mu\}$ and $\mu^\downarrow := \{\mu' \in \W: \mu \rhd \mu'\}$. Second, for any two matchings, $\mu,\mu'$, and a set $T \subset \W$ such that $\mu,\mu'\in T$, let $\AthxyA := \{ \{\mu,\mu',\hat \mu\} : \hat \mu \in T\}$. That is, $\AthxyA$ is a collection of all possible subsets of $T$ of size at most $3$ that contain $\mu$ and $\mu'$.  Notice that by having $\hat \mu = \mu$ or $\mu'$, $\{\mu,\mu'\} \in \AthxyA$. 

Now, I present three normatively desirable axioms for a refinement. The main result of this section, Theorem \ref{Theorem: Characterization of C}, shows that these axioms characterize $\C$. 
\smallskip

\begin{itemize}
\item[] \textbf{Include the Maximal Element (IM):} 
\begin{quote} For any $T \subset \W$ and $\mu \in T$, if $\mu^\uparrow \cap T =\emptyset$, 
then $\mu \in \G(T)$. Moreover, if $\rhd$ is a strict and a complete order on $T$, then $\G(T) = \{\mu \in T: \mu \rhd \mu' \; \forall \mu' \in T\}.$
\end{quote}

This axiom requires that, if $\mu$ is a maximal element in $T$, then $\G(T)$ must include $\mu$, and if $T$ is totally ordered by $\rhd$, then $\G(T)$ must select only the unique such element.

\item[] \textbf{Exclude the bottom (EB):} 

\begin{quote}For any $T \subset \W$, if $\mu^\downarrow \cap T = \emptyset$ and $\mu^\uparrow \cap T \neq \emptyset$, then $\mu \notin \G(T)$. \end{quote}

This axiom requires a refinement to exclude the ``bottom layer'' according to $\rhd$. That is, if a matching $\mu$ does not dominate any other matching in $\W$, and is also strictly dominated by some other matching $\mu' \in \W$, then a good refinement should exclude $\mu$ according to EB.

\item[] \textbf{Expansion from triples (ET):}

\begin{quote}If, for some $\mu,\mu',T$ such that $\mu,\mu'\in T \subset \W$, $\mu' \notin \G(T')$ $\forall T' \in \AthxyA$, then $\mu' \notin \G(T)$. \end{quote}

This axiom is a form of consistency requirement from observing $\G(\cdot)$'s behaviour on sets of size $3$ or less. Alternatively, it can also be loosely interpreted as a form of revealed preference. The axiom says the following. Fix a set $T$ that contains $\mu$ and $\mu'$. If $\mu'$ is never chosen from any subset of $T$ of size $3$ or less that includes both $\mu$ and $\mu'$, then $\mu'$ should not be chosen from $T$. 
\end{itemize}
While the IM and EB axioms seem relatively natural, the ET axiom perhaps  merits some discussion. This axiom above is, essentially, a weakening of WARP. Similar weakenings of WARP have appeared in the literature on incomplete preferences in choice theory, e.g., \cite{eliaz2006indifference}. More closely, this axiom is similar to the ``upward consistency'' axiom from \cite{nishimura2018transitive} that deals with inferring a transitive part of a possibly cyclic choice behavior. 

That the above three axioms are mutually independent is straightforward to see. In light of that observation, I present the main result of this section---a characterization of $\C$ using IM, EB and ET.

\bthm\label{Theorem: Characterization of C} Any refinement $\G(\cdot): 2^\W \to 2^\W$ that satisfies IM, EB and ET has $\G(\cdot) \subseteq \C(\cdot)$. Conversely, $\C(\cdot)$ satisfies IM, EB and ET. Hence, $\C(\cdot)$ is the most permissive refinement of $\W$ satisfying IM, EB and ET. \ethm 


\subsection{Relation to other solution concepts}
The characterization in Theorem \ref{Theorem: Characterization of C} leaves open the possibility that the compromise core can contain matchings $\mu$ and $\mu'$ such that $\mu \rhd \mu'$. This may be unsatisfactory if one holds the perspective that a solution concept describes a ``system at rest.'' That is, we may want a solution concept to not include two matchings wherein one dominates the other. The vNM stable set (\cite{von1944morgenstern}) is one such concept. In our context, one would say that a set $S$ is a vNM stable set if it satisfies the following two properties. 

\begin{itemize}
\item[] \textit{Internal stability:} No $\mu, \mu' \in S$ has $\mu \rhd \mu'$. 
\item[] \textit{External stability:} If $\mu \notin S$, then some $\mu' \in S$ has $\mu' \rhd \mu$. 
\end{itemize}
The above two axioms, while seemingly natural, can easily lead to the situations where no vNM stable set exists. For example, consider Example \ref{Example: empty the strong core}. There, $\W = \{ \mu_1, \mu_2, \mu_3\}$ with  $\mu_1 \rhd \mu_2 \rhd \mu_3 \rhd \mu_1$. Internal stability demands that any candidate stable set can have at most one element. However, no such set can satisfy external stability. 

Another solution concept popularly used in cooperative games is the bargaining set.\footnote{For an example of the bargaining set in the matching environment, see \cite{atay2021bargaining}.} Loosely speaking, if $\mu' \rhd \mu$ due to a move enforced by $S$, then we say that $(S,\mu')$ is an objection to $\mu$. However, if there is a set $T$ such that some (but not all) members of $S$ are in $T$ and $(T,\mu'')$ is an objection to $\mu'$, then we say that $(T,\mu'')$ is a counterobjection against an objection $(S,\mu')$. A matching $\mu$ is in the bargaining set if, for every objection there is a counterobjection. 

Regardless of whether the bargaining set is always nonempty (which I conjecture it is), or whether a vNM stable set exists, they are unappealing because they can fail a natural requirement of EB: Exclude the bottom. Consider Example \ref{Example: problem with Fishburn covering} given in the Appendix.\footnote{I have avoided presenting the full example here primarily because I only use its weak core for the present argument.} Its weak core contains three elements, $\{\mu_1,\mu_2,\mu_3\}$ such that $\mu_2\rhd \mu_1$, $\mu_3 \rhd \mu_2$, but $\mu_3$ and $\mu_1$ are not $\rhd$-comparable. What would we want as a reasonable solution concept in this example? First, it is uncontentious that it should include $\mu_3$, an $\rhd$-undominated matching. Should it include $\mu_1$? For one, $\mu_1$ is $\rhd$-dominated by $\mu_2$, and it does not dominate any other matching. Therefore, if a refinement needs to include $\mu_1$, we may do better by including $\mu_2$ and excluding $\mu_1$ instead. 

Indeed, this is precisely the compromise core, i.e., $\C = \{\mu_2,\mu_3\}$. On the other hand, the only vNM stable set and the bargaining set are $\{\mu_1,\mu_3\}$. By including $\mu_1$ and excluding $\mu_2$, the vNM stable set and the bargaining set not only include a dominated matching which does not dominate any other matchings, but also exclude its upper contour set.\footnote{An upper contour set of a matching $\mu$ is $\mu^\uparrow$.} 

It is a worthwhile goal to refine $\C$ further, given that it is the most permissive solution concept satisfying the three natural normative properties mentioned before Theorem \ref{Theorem: Characterization of C}. A straightforward way to do that, while achieving internal stability, would be to select the largest internally stable subsets of $\C$. Of course, any such set may not satisfy external stability that $\C$ does, as Proposition \ref{Proposition: C external stability} below shows.
\bprop\label{Proposition: C external stability} If $\mu \in \W \backslash \C$, then $\exists \mu' \in \C$ such that $\mu' \rhd \mu$. \eprop

\section{Men- and Women-Optimal Core}\label{Section: men and women optimal core}
We now turn our attention to another object of interest for applications---``best'' stable matchings according to one side of the market. For example, in school choice, it is reasonable to favor students and select a stable matching that they prefer the most. As is standard, I define an order $\succ_M$ on $\W$ to mean, $\mu \succ_M \mu'$ iff there is no $i \in M$ such that $\mu'(i) \succ_i \mu(i)$ \emph{and} for at least one $i \in M$, $\mu(i) \succ_i \mu'(i)$. Order according to women, $\succ_W$, is defined analogously.\footnote{This definition is, unfortunately, slightly different from $\succ_S$ defined in Definition \ref{Definition: succ_S order}. In particular, we do not require $\mu(M) = M$ here. Instead of choosing a different notation, I still choose to denote it by $\succ_M$ for ease of exposition. }

When the preferences are complete, strict and transitive, the set of stable matchings is a  lattice under the order $\succ_M$ or $\succ_W$. Therefore, there is a unique men- (women-) optimal stable matching that is the maximum element of the set of stable matchings according to $\succ_M (\succ_W)$. With weak \emph{but transitive} (and complete) preferences, a maximum element need not exist. A natural remedy is to weaken the definition to \emph{maximal} elements---i.e., say that $\mu$ is men-optimal if there is no  weakly stable $\mu'$ such that $\mu' \succ_M \mu$. As mentioned before, \cite{erdil2017two} show that such a matching exists. Unfortunately, when preferences are incomplete, matchings that satisfy even this weaker definition may not exist. The reason for this lack of existence of a maximal element is, unsurprisingly, the intransitivity of $\succ_M$.   

This intransitivity motivates an obvious definition of men- and women-optimal core analogous to the compromise core. 

\bdefn\label{Definition: man and woman optimal core} $\mu$ M-covers $\mu'$, denoted by $\mu \ll^M \mu'$ if $\mu \succ_M \mu'$ and $\mu' \succ_M \mu''$ $ \implies \mu \succ_M \mu''.$ The men-optimal core, $\W^M$, is defined as $\text{MAX}(\C, \ll^M)$.  Women optimal core, $\W^W$, is defined analogously as the set of $\ll^W$-uncovered elements. \edefn
\brem Notice that the men-optimal core selects the $\ll^M$-maximal elements from $\C$ and not $\W$. One could, alternatively, define the men-optimal core to be the $\ll^M$-maximal elements of $\W$. Unsurprisingly, the two notions need not coincide. I take $\C$ as a primitive primarily to ensure that the men-optimal core is a subset of $\C$. \erem 

In applications such as school choice, when students have incomplete preferences, forcing them to report complete preferences could yield outcomes that are Pareto dominated for the students. 
This issue is not new: Similar problems arise if we replace incompleteness with indifference. However, so long as the weak preferences are transitive one can use the \cite{erdil2017two}'s algorithm to obtain a men-optimal stable matching. In contrast, reinterpreting incompleteness as indifference may result in intransitivities, rendering this algorithm unusable. 

Let us consider the IIT-JEE admissions process mentioned in the Introduction. In the current system, students are ranked according to their performance on an exam. Therefore, each object (seat at one of the IITs) ranks the students the same way. Each student submits a strict ranking over the set of objects---i.e., institute and undergraduate major. Then, a version of the student-proposing deferred acceptance algorithm (DAA henceforth) is executed to achieve a matching.\footnote{\cite{baswana2015joint} provide precise details of the procedure.} 
As the following example shows, by forcing agents to report complete preferences and running a student-proposing DAA, we may obtain a strictly suboptimal matching for the students.

\bexam\label{Example: application students engineering first example}Let $W=\{w_1,w_2,w_3\}, M = \{m_1, m_2,m_3\}$. Men rank women as $w_1 \succ_m w_2 \succ_m w_3$. Women's preferences are: $m_1 \otimes_w m_2, m_1 \otimes_w m_3, m_2 \succ_w m_3$ for all $w\in W$. Finally, all the agents strictly prefer being matched over remaining single. 
\eexam

It is straightforward to check that $\W = \{\mu_1, \mu_2, \mu_3\}$, where
\begin{align*}
\mu_1 = \{ (w_1, m_1), (w_2,m_2), (w_3,m_3)\}\\
\mu_2 = \{ (w_1, m_2), (w_2,m_1), (w_3,m_3)\}\\
\mu_3 = \{ (w_1, m_2), (w_2,m_3), (w_3,m_1)\}
\end{align*}
First, notice that $\C = \W$. Moreover, $\mu_1 \succ_W \mu_3$ while $\mu_2$ and $\mu_3$ or $\mu_1$ and $\mu_2$ are not $\succ_W$ comparable. Therefore, the set of women-optimal stable matchings---the set of $\ll^W$ uncovered elements in $\C$---is $\{\mu_1, \mu_2\}$. 

However, notice that forcing women to submit a complete ranking and running a DAA thereafter can generate all three matchings in $\C$. In particular, if the women submit the following preference: $m_2 \succ_w m_3 \succ_w m_1$---a completion of their original preference---then the outcome would be $\mu_3$, a strictly inferior outcome for women compared to $\mu_1$.

If we had allowed the women to submit incomplete preferences, 
we could have chosen between $\mu_1$ and $\mu_2$. Therefore, allowing students to express incomplete preferences may generate welfare gains. These could, in principle, be substantial given the size of the market in applications such as college admissions. For example, the number of students that gained admissions just through the single IIT-JEE seat-allocation mechanism was 13,500 in 2019, and 300 seats ($2\%$ of the total capacity) went vacant.\footnote{{https://www.hindustantimes.com/mumbai-news/over-300-iit-seats-allotted-in-final-round-of-admissions/story-VTKpYlsDxQ9TRpKOiW48NJ.html.}} 

\section{Related Literature}

This paper contributes to the literature on matching with nonstandard preferences.
A few papers study matching problems with incomplete preferences from a theoretical point of view.
\cite{manlove2002structure} studies the lattice structure of the stable matchings for three different notions of stability, two of which are the weak and strong stability considered in this paper.
\cite{irving1994stable} provides algorithms for weak and strong stable matchings allowing incomplete preferences.
There is a recent and burgeoning literature studying matchings with incomplete preferences from a computational point of view.\footnote{See for example \cite{manlove2002hard}, \cite{irving2003strong}, \cite{aziz2017stable} and \cite{cseh2018pairwise}. }
For example, \cite{bade2016pareto} studies optimality of matching algorithms when preferences are modeled by choice functions. 
\cite{chambers2017choice} develop a deferred acceptance algorithm for matchings when preferences are given as choice correspondences.\footnote{Also see \cite{hatfield2005matching} and \cite{aygun2013matching} for more seminal work on matching where preferences are modeled by choice functions.}
\cite{che2019weak} show an existence theorem of many to one stable matchings for general preferences that allow for incompleteness. In particular, they give a fixed-point characterization of the strong core when one side of the market can have incomplete preferences. In a contemporaneous paper, \cite{kitahara2021stable} study school choice problems allowing for schools to find some students incomparable. They propose strategy proof mechanisms to obtain weakly stable matchings. In contrast, my focus is to offer a new solution concept---the compromise core. In matching with externalities, an agent's preference over two outcomes may be incomplete \emph{until an agent knows how the rest of the market is organized.} Several papers have studies such markets in the context of various applications, e.g. \cite{sasaki1996two}, \cite{dutta1997stability}, \cite{echenique2007solution}, \cite{pycia2012stability}, \cite{pycia2021matching}. Recently, \cite{caspari2021non} study a classical two-sided matching model when agents may exhibit nonstandard choice behaviour. They provide conditions under which a stable matching exists and also when a strategyproof mechanism exists to achieve a stable matching. 

While all of the above papers study properties of existing solution concepts under nonstandard preferences, including questions related to their existence, my focus is on proposing new solution concepts suitable for matching problems when agents have incomplete preferences. 

Preference incompleteness has been widely studied in decision theory and behavioural economics (e.g. \cite{ok2002utility}, \cite{eliaz2006indifference}), and its implications have also been studied in strategic environments, e.g., \cite{bade2016pareto}. Incompleteness of preferences is closely related to indifference.
Indeed, most of the findings in this paper can be viewed as results for matchings with intransitive indifference.
As such, I encounter problems arising from indifference.
For example, under indifference, stable matchings may not be efficient.
\cite{erdil2017two} study this issue and propose an algorithm to obtain efficient matchings. 

In highlighting the importance of allowing the agents to express indifferences, \cite{erdil2017two} say, ``\emph{...allowing, even encouraging, agents to express indifferences when ranking alternatives not only simplifies preference revelation and market participation, but also improves efficiency}''. They argue that indifference has importance beyond theoretical interest.\footnote{For example, \cite{abdulkadirouglu2009strategy} show empirically that the tie-breaking rule in a school choice mehanism has significant welfare implications.}
Incompleteness exacerbates these concerns as efficient matchings may not even exist.
I have proposed alternative notions of stability that seem to be more suitable for such environments.
%

Lastly, the definition of the compromise core draws on a rich tradition of uncovered sets in social choice theory. The basic principle therein (adopted in this paper) is to construct a transitive order using the covering relation using the underlying intransitive order. Besides the Miller covering (used in this paper), there are several other notions of covering, e.g \cite{gillies1959solutions}, \cite{fishburn1977condorcet}, \cite{bordes1983possibility}, \cite{mckelvey1986covering}.\footnote{These, and many others, are extensively studied in \cite{duggan2013uncovered}.}  \cite{fishburn1977condorcet} has also been found useful in problems of choice theory, e.g. \cite{nishimura2018transitive}. In fact, \cite{dutta1999comparison} offer a more general treatment of the problem of constructing a transitive order on a choice set that goes beyond covering using ``comparison functions'' that have the uncovered sets as a special case.

\section{Conclusion}\label{Section: Conclusion}
I study a two-sided matching market in which agents may have incomplete preferences. Incompleteness is pervasive and, especially in the context of matching, it has important consequences for stability. Indeed, choice-theoretically, incompleteness is indistinguishable from intransitive indifference. I prefer not to take a strong stand on which interpretation is more reasonable, as the main contribution in this paper---formulation of the compromise core and the men- and women-optimal core---is equally valid under either interpretation.

The goal of this paper has been largely exploratory, in the sense that I have attempted to study the effects of agents having incomplete preferences on the usual notions of stability. I hope that concepts such as the compromise core offer a good candidate for stability that is more permissive than the strong core and more restrictive than the weak core. At the same time, should one find my proposals reasonable, a natural next step is to seek efficient algorithms that produce a matching in the compromise core, or in the men-optimal core. When the strong core is non-empty, \cite{irving1994stable} provides such an algorithm. Therefore, it is natural to explore such algorithms when the strong core is empty. 

Equally importantly, I have abstracted away from strategic considerations. Again, from a practical standpoint, these are of paramount importance. Therefore, another worthwhile goal would be to explore strategy-proof mechanisms (for the proposing side), such as the DAA, which produce, say, matchings in the men-optimal core.

\bibliography{ref-matching}

\newpage

\appendix

\section{Appendix}

\bprf[Proof of Proposition \ref{Proposition: equivalence of core and stability}] If $\mu$ is in the weak core, it is obviously weakly stable. For the opposite, suppose $\mu$ is weakly stable but is not in the weak core. Therefore,  there is $S \subset A$ and a matching $\mu'$ such that $\mu'(i) \in S$ for all $i \in S$ and, $\mu'(i) \succ_i \mu(i)$ $\forall i \in S$. Pick any $i \in S$ and let $j:=\mu'(i)$.
Then $ j\succ_i \mu(i)$ and $i \succ_j \mu(j)\neq i$. Therefore, $(i,j)$ is a strong blocking pair for $\mu$, a contradiction. Therefore, $\mu$ is in the weak core. 

Similarly, if $\mu$ is in the strong core then it is strongly stable. For the reverse, suppose $\mu$ is strongly stable but there is $S \subset A$ and a $\mu'$ such that $\mu'(i) \in S$ for all $i\in S$ and, either $\mu'(i) \succ_i \mu(i)$ or $\mu'(i) \otimes_i \mu(i)$ for all $i\in S$. Moreover, at least for one $i\in S$, $\mu'(i) \succ_i \mu(i)$. Let $i^*$ be such an agent. But then, $(i^*, \mu'(i^*))$ is a weak blocking pair, a contradiction. Therefore, $\mu$ is in the strong core. 
\eprf

\bprf[Proof of Proposition \ref{Proposition: weak stable iff stable according to some completion}]
\noindent Suppose $\mu$ is stable according to some completion $\comp$. Notice that any blocking pair in $\succsim$ remains to be a blocking pair according to $\comp$. Therefore, $\mu$ is weakly stable according to $\succsim$.

\noindent For the converse, suppose $\mu$ is weakly stable. 
Let $C(\succsim)$ denote the set of all the completions of $\succsim$. Define, $$T:= \{(m,w): (m,w) \text{ is a blocking pair for } \mu \text{ for some }\comp \in C(\succsim)\}.$$ 

We now prove a claim towards completing the proof. 
\bclaim\label{Claim: if a blocking pair then one is incomparable} If $(m,w) \in T$, then either $w \otimes_m \mu(w)$ or $m\otimes_w \mu(w)$, or both, according to $\succsim$. \eclaim 
\bprf 
Suppose not. So, $\exists (m,w) \in T$ such that $m$ finds $w$ and $\mu(m)$ comparable, and $w$ finds $m$ and $\mu(w)$ comparable. Since it is a blocking pair for some $\comp$, it must be the case that $w \comp_m \mu(m)$ and $m \comp_w \mu(w)$. Since $w$ and $\mu(m)$ are comparable according to $\succsim_m$, we have $w\succsim_m \mu(m)$. Similarly, $m\succsim_w \mu(w)$. Therefore, $(m,w)$ is a strong blocking pair for $\mu$ according to $\succsim$, i.e. $\mu$ is not weakly stable. A contradiction. 
\eprf
Now, for each $i \in T$, define $S(i) := \{j: (i,j) \in T \text{ or } (j,i) \in T \text{ and } j \otimes_i \mu(i)\}$.\footnote{The distinction between $(i,j)$ and $(j,i)$ is merely because we have defined a blocking pair as an ordered pair $(m,w)$. Therefore, whether $i \in M$ or $i\in W$ would matter for notational consistency. }  
Consider a completion $\comp^*$ of $\succsim$ such that, for all $i \in A$, $\mu(i) \comp^*_i j$ for all $j \in S(i)$.\footnote{That such a completion exists is a straightforward consequence of the Szpilrajn extension theorem. We define a new partial order, $\widehat \succ_i$ in two steps. First, for any $j$ such that $\mu(i) \otimes_i j$, we set $\mu(i) \widehat \succ_i j$. Then, we take its transitive closure to make $\widehat \succ_i$ a partial order. Finally, by using the Szpilrajn extension theorem, we extend this partial order to a complete linear order.}

We complete the proof by showing that $\mu$ is stable according to $\comp^*$. 
Suppose not. $\exists$ a blocking pair $(m,w)$ according to $\comp^*$. But then, $w \in S(m)$ or $m \in S(w)$ by Claim \ref{Claim: if a blocking pair then one is incomparable}. Suppose, wlog, $w \in S(m)$. But then, by construction, $\mu(w) \comp^*_w m$. Therefore, $(m,w)$ cannot be a blocking pair for $\mu$ according to $\comp^*$. A contradiction. Hence, $\mu$ is stable according to $\comp^*$. 
\eprf

\bprf[Proof of Proposition \ref{Proposition: strong core not intersection}]
Suppose that $\mu$ is stable according to every completion of $\succ$ but is not strongly stable. That means, wlog, there is a blocking pair $(m,w)$, such that $m \succ_w \mu(w)$ and $w \otimes_m \mu(m)$. Consider a completion, $\comp$, of $\succ$ such that $w \succ_m \mu(m)$. Such a completion exists by the Szpilrajn theorem as in the proof of Proposition \ref{Proposition: weak stable iff stable according to some completion}. Therefore, $(m,w)$ is a blocking pair in $\mu$ in $\comp$, i.e., $\mu$ is not stable in $\comp$. A contradiction. Therefore, $\mu$ must be strongly stable.

\eprf

\blemma \label{Lemma: lhd incompelte, intransitive, antisymmetric} $\rhd$ is a possibly incomplete and \emph{intransitive} binary relation over $\mc M$. Also, it is possible to have $\mu \rhd \mu' \rhd \mu$. 
\elemma

\bprf
That $\rhd$ is possibly incomplete is obvious. Consider $M = \{m_1, m_2\}, W = \{w_1,w_2\}$. Suppose that all the agents strictly prefer being matched over remaining single. Moreover, $w_1 \succ_{m_1} w_2$,$w_2 \succ_{m_2} w_1$, $m_2 \succ_{w_1} m_1$ and $m_1 \succ_{w_2} m_2$. It is easy to check $\mu = \{(m_1,w_1), (m_2,w_2)\}$ and $\mu' = \{(m_1,w_2), (m_2, w_1)\}$ are in $\W$ and are not ranked according to $\rhd$. 

The intransitivity of $\rhd$ is seen in a number of examples throughout the paper, one being Example \ref{Example intro: nontransitivity of strong dominance}. 

\eprf 

\blemma\label{Lemma: covering is transitive}$\ll$ is antisymmetric and transitive. \elemma
\bprf For transitivity, suppose that $\mu \ll \mu'$ and $\mu' \ll \mu''$. Consider any matching $\hat \mu$ such that $\mu'' \rhd \hat \mu$. Since $\mu' \ll \mu''$, we have $\mu' \rhd \hat \mu$. Since $\mu \ll \mu'$, $\mu \rhd \hat \mu$. Therefore, $\mu \ll \mu''$. 

For its antisymmetry, suppose that $\mu \ll \mu' \ll \mu$. Therefore, $\mu' \rhd \mu$. Since $\mu \ll \mu'$, we have $\mu \rhd \mu$. Since $\rhd$ is not-reflexive, this is not possible. 
\eprf

\bprf[Proof of Proposition \ref{Proposition: strong dominance iff the strong core}] 
Suppose $\mu \in \cs$ and there is a $\mu'$ such that $\mu' \rhd \mu$. Therefore, there exists an $S \subset A$ such that $S$ can enforce $\mu'$ over $\mu$ and $\mu' \succ_S \mu$. Hence, there is $i \in S$ such that $\mu' \succ_i \mu$. Moreover, for the partner of $i$ in $\mu'$, denoted by $j:= \mu'(i)$, either $i \succ_j \mu(j)$ or $i \otimes_j \mu(j)$. But then, $(i,j)$ is a weak blocking pair for $\mu$, i.e. $\mu \notin \cs$, a contradiction. 

\eprf

\bprf[Proof of Theorem \ref{Theorem: Characterization of C}]
That $\C$ satisfies IM and EB is obvious. 
\bclaim $\C(\cdot)$ satisfies $ET$. \eclaim 
\bprf Suppose that, for some $\mu,\mu' \in A \subset \W$, $\mu' \notin \C(A')$ $\forall A' \in \AthxyA$ but $\mu' \in \C(A)$. Therefore, $\mu'$ is uncovered in $A$. First, suppose that $\mu \rhd \mu'$. Since $\mu'$ is uncovered in $A$, $\exists \hat \mu \in A$ such that $\mu' \rhd \hat \mu$ but not $\mu \rhd \hat \mu$. But then, $\mu' \in \C( \{\mu,\mu',\hat \mu\})$. Since $\{\mu,\mu',\hat \mu\} \in \AthxyA,$ this is a contradiction. Therefore, $\mu \rhd \mu'$ is not possible. On the other hand, if $\mu' \unrhd \mu$ or if $\mu$ and $\mu'$ are not $\rhd$ ranked, then $\mu' \in \C(\{\mu,\mu'\})$: a contradiction as $\{\mu,\mu'\} \in \AthxyA$. 
\eprf

For the reverse, suppose $\G$ satisfies the IM, EB and ET. Let $\ll^A$ denote the covering relation on a set $A$. That is, for any $\mu,\mu' \in A$, $\mu \ll^A \mu'$ if $\mu \rhd \mu'$ and $\mu' \rhd \hat \mu \implies \mu \rhd \hat \mu$ for all $\hat \mu \in A$. 

\bclaim $\mu \ll^A \mu' \implies \mu' \notin \G(A)$. \eclaim 

\bprf Since $\mu \ll^A \mu'$, we have that $\mu \rhd \mu'$. Therefore, if $A = \{\mu,\mu'\}$, $\mu' \notin \G(A)$ due to IM. Suppose $A \neq \{\mu,\mu'\}$. 
Consider any $\hat \mu \in A$ and let $A':=\{\mu,\mu',\hat \mu\}$. We now prove that $\mu' \notin \G(A')$. 

First, suppose that $\mu' \rhd \hat \mu$. 
Notice that, $\mu \ll^A \mu' \implies \mu \rhd \hat \mu$. Therefore, $A'$ is totally ordered, and hence, by IM, $\G(A') = \{\mu\}$.

Alternatively, if $\mu' \unrhd \hat \mu$ (and not $\mu' \rhd \hat \mu$) or if $\mu'$ and $\hat \mu$ are not $\unrhd$-ranked, then $\mu'^\uparrow \cap A' \ni \mu$ and $\mu'^\downarrow \cap A' = \emptyset$. Therefore, $\mu' \notin \G(A')$ by EB. 

Therefore, $\mu' \notin \G(A')$. However, if $\mu' \notin \G(\{\mu,\mu',\hat \mu\})$ for any $\hat \mu \in A$, then $\mu' \notin \G(A)$ by ET. 
\eprf 

That is, if $\mu'$ is covered in $A$, then $\mu' \notin \G(A)$. In other words, $\G(A) \subset \C(A)$ $\forall A \subset \W$.
\eprf

\bprf[Proof of Proposition \ref{Proposition: C external stability}]
Suppose that $\mu \in \W \backslash \C$. Therefore, $\exists \mu_1 \ll \mu$. If $\mu_1 \in \C$ we are done. If not, $\exists \mu_2 \ll \mu_1 \implies \mu_2 \rhd \mu$ (since $\mu_1 \rhd \mu$). Since $\ll$ is transitive and strict and $\W$ is finite, $\exists \mu_n$ such that $\mu_n$ is not covered by any other matching and $\mu_n \rhd \mu$. Since $\mu_n$ is uncovered, $\mu_n \in \C$ establishing the proposition. 
\eprf

\bexam\label{Example: problem with Fishburn covering}
Let $M = \{m_1,m_2,m_3\}$ and $W= \{w_1,w_2,w_3\}$. Men's preferences are the following: $w_2 \succ_{m_1} w_1 \succ_{m_1} w_3$, $w_2 \succ_{m_2} w_3, w_1 \otimes_{m_2} w_2,w_3$, $w_1 \succ_{m_3} w_3, w_2 \otimes_{m_3} w_1,w_3$. All the men strictly prefer being matched over remaining single. Women's preferences are the following: $m_1 \succ_{w_1} m_3$ and $m_2 \otimes_{w_1} m_1,m_3$. Also, $w_1$ strictly prefers being matched over remaining single. For $w_2$, being matched to $m_1$ or $m_2$ is strictly preferred to remaining single. But, being single is strictly preferred over being matched to $m_3$. She cannot compare $m_1$ and $m_2$. Similarly, $w_3$ strictly prefers being matched to $m_2$ or $m_3$ over remaining single, and prefers remaining single over being matched to to $m_1$. She cannot compare $m_2$ and $m_3$. \eexam

\bclaimst $\W = \{\mu_1, \mu_2, \mu_3\}$ where $\mu_1 = \{(m_1,w_1), (m_2, w_2), (m_3,w_3)\}$, \\$\mu_2 = \{(m_1,w_2), (m_2,w_1), (m_3,w_3)\}$ and $\mu_3 = \{(m_1,w_2),(m_2,w_3),(m_3,w_1)\}$.\eclaimst

\bprf 
It is easy to check that $\mu_1, \mu_2$ and $\mu_3$ are weakly stable. To rule out other matchings, first notice that matchings in which $w_3$ is matched to $m_1$ or $m_3$ is matched to $w_2$ are not in the weak core. That rules out the remaining matches wherein all the agents are matched. Second, if there is a matching in $\W$ wherein one man-woman pair is single it can either involve $w_2$ and $m_3$ being single or $w_3$ and $m_1$ being single. Suppose $w_2$ and $m_3$ are single. Then, either we have $\{(m_1,w_1),(m_2,w_3)\}$ or $\{(m_1,w_3), (m_2,w_1)\}$ as the possible matches. In the first matching, $(m_2,w_2)$ constitute a strong blocking pair while in the second, $(m_1,w_3)$ is ruled out as $w_3$ prefers remaining single over being matched to $m_1$. Similarly, it is easy to check that matchings wherein $m_1$ and $w_3$ are weakly unstable too. Lastly, it is obvious that there cannot be a weakly stable matching with two or more men unmatched. 

\eprf 
\bclaimst The Fishburn uncovered set in Example \ref{Example: problem with Fishburn covering} is $\{\mu_1, \mu_3\}$ while $\C = \{\mu_2, \mu_3\}$. \eclaimst
\bprf 
It is easy to check that $\mu_2 \rhd \mu_1$ and $\mu_3 \rhd \mu_2$. However, $\mu_1$ and $\mu_3$ are not $\rhd$ comparable. Therefore, $\C = \{\mu_3, \mu_2\}$ as $\mu_2 \rhd \mu_1$ but not $\mu_3 \rhd \mu_1$. 

However, if we used Fishburn covering instead of Miller covering, then $\mu_2$ is covered by $\mu_3$. Recall that $x$ F-covers $y$ if, $z \rhd x \implies z \rhd y$. Since no matching $\rhd$-dominates $\mu_3$, $\mu_3$ F-covers $\mu_2$ trivially. On the other hand, $\mu_1$ is F-uncovered. Therefore, the Fishburn uncovered set here is $\{\mu_1, \mu_3\}$. 
\eprf 

\subsection{Example illustrating the compromise core}\label{Section: appendix example of a compromise core}

\begin{example}\label{Example: motivation IC}
Let $M = \{m_1,m_2,m_3,m_4\}$ and $W= \{w_1,w_2,w_3,w_4\}$. Consider the preferences of men given by the Hasse diagram in Figure \ref{fig:motivation IC_2}. 
\begin{figure}[h]
\begin{center}
\includegraphics[width=14cm]{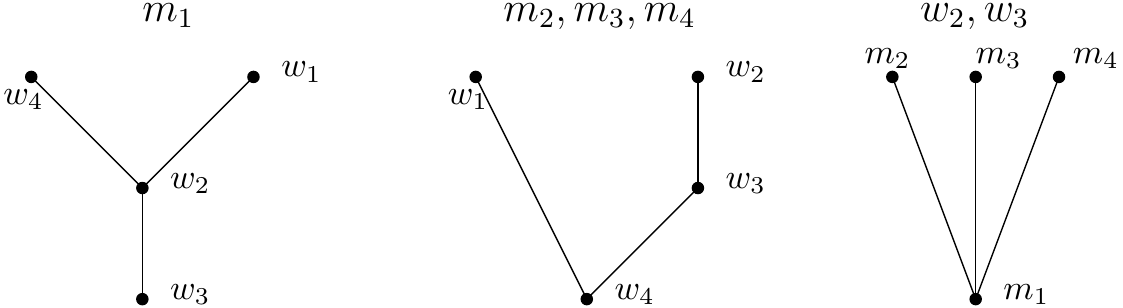}
\caption{Preferences represented via Hasse diagram}
\label{fig:motivation IC_2}
\end{center}
\end{figure}

Moreover, $w_1$ and $w_4$ strictly prefer being matched to remaining single, but find every man incomparable. 
\end{example}

The Hasse diagram offers a succinct way of representing preferences. For the reader unfamiliar with it, here are the preferences. For $m_1$, $w_1, w_4 \succ_{m_1} w_2 \succ_{m_1}w_3$ and $w_1 \otimes_{m_1} w_4$. For $m \in \{m_2,m_3, m_4\}$, we have, $w_2 \succ_m w_3 \succ_m w_4$, $w_1 \succ_m w_4$, $w_1 \otimes_m w_2$ and $w_1 \otimes_m w_3$. 
For $w \in \{w_2,w_3\}$, $m_2,m_3,m_4 \succ_w m_1$ and $m_i \otimes_w m_j$ if $i\neq j$ and $i,j \in \{2,3,4\}$. 
For $w \in \{w_1,w_4\}$, $m_i \otimes_{w} m_j$ iff $i\neq j$ and $i,j\in \{1,2,3,4\}$, and $m_i \succ_w w$ for all $m_i \in M$.

It is straightforward to see that the strong core is empty, and the weak core contains all possible matchings such that $\mu(m_1) \in \{w_1, w_4\}$; and no agent is unmatched. However, there is a sense any matching $\mu \in \W$, such that $\mu(m_1) = w_4$ are better than those where $\mu(m_1) = w_1$. Indeed, both of these types of matchings are in the weak core. However, when $\mu(m_1) = w_1$, $\mu(\{m_2,m_3,m_4\}) = \{w_2,w_3,w_4\}$.\footnote{Recall that $\mu(S):= \{j : j = \mu(i) \text{ for some }  i\in S \}$.} On the other hand, when $\mu(m_1) = w_4$, $\mu(\{m_2,m_3,m_4\}) = \{w_1,w_2,w_3\}$. 
Therefore, given any matching in the weak core that matches $m_1$ with $w_1$, all the agents could decide to move to another matching in the weak core, where $m_1$ is matched with $w_4$. This way, it is possible to make some agent strictly better off without making any agent worse off. More importantly, a similar move in the opposite direction---from $m_1$ being matched with $w_4$ to $m_1$ being matched with $w_1$---would always make some agent strictly worse off. We now see that the compromise core would select precisely those matchings from the weak core where $\mu(m_1) = w_4$. 

\blemma In Example \ref{Example: motivation IC}, $\C = \{\mu \in \W : \mu(m_1) = w_4\}$.\elemma 

\bprf 

Below, we enumerate all the matchings in $\W$, such that $\mu(m_1) = w_1$.
\begin{eqnarray*}
\mu_1 &= \{(m_1,w_1), (m_2,w_2),(m_3,w_3), (m_4,w_4)\}\\
\mu_2 &= \{(m_1,w_1), (m_2,w_2),(m_3,w_4), (m_4,w_3)\}\\
\mu_3 &= \{(m_1,w_1), (m_2,w_3),(m_3,w_2), (m_4,w_4)\}\\
\mu_4 &=\{(m_1,w_1), (m_2,w_3),(m_3,w_4), (m_4,w_2)\}\\
\mu_5 &= \{(m_1,w_1), (m_2,w_4),(m_3,w_2), (m_4,w_3)\}\\
\mu_6 &= \{(m_1,w_1), (m_2,w_4),(m_3,w_3), (m_4,w_2)\}
\end{eqnarray*}
Now, we enumerate all the matchings in $\W$, such that $\mu(m_1) = w_4$. 
\begin{eqnarray*}
\mu_7 &= \{(m_1,w_4), (m_2,w_2),(m_3,w_3), (m_4,w_1)\}\\
\mu_8 &= \{(m_1,w_4), (m_2,w_2),(m_3,w_1), (m_4,w_3)\}\\
\mu_9 &= \{(m_1,w_4), (m_2,w_3),(m_3,w_2), (m_4,w_1)\}\\
\mu_{10} &=\{(m_1,w_4), (m_2,w_3),(m_3,w_1), (m_4,w_2)\}\\
\mu_{11} &= \{(m_1,w_4), (m_2,w_1),(m_3,w_2), (m_4,w_3)\}\\
\mu_{12} &= \{(m_1,w_4), (m_2,w_1),(m_3,w_3), (m_4,w_2)\}
\end{eqnarray*}

\begin{figure}[h!]
\begin{center}
\includegraphics[width=11cm]{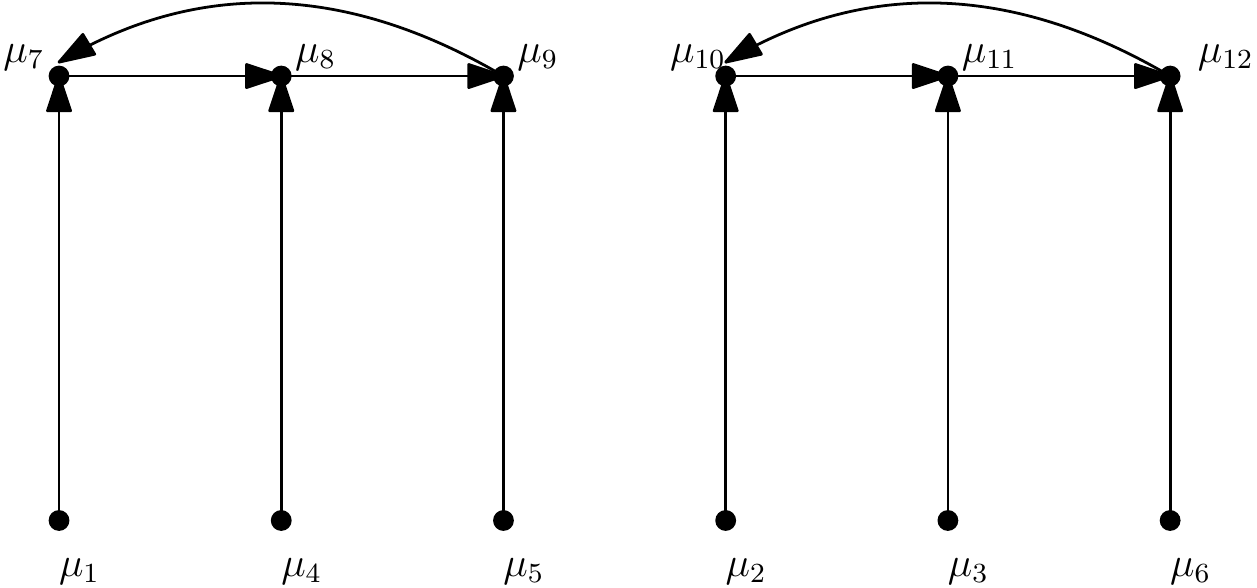}
\caption{Graph with nodes as matchings in $\W$.}
\label{fig:IC graph weak core motivating example}
\end{center}
\end{figure}

Figure \ref{fig:IC graph weak core motivating example} depicts the relation $\unrhd$ in a graph with each node being a matching in $\W$.\footnote{To avoid cluttering the graph, I have chosen to avoid some edges---for example, an edge from $\mu_1$ to $\mu_{11}$. More importantly, there is no edge from $\mu_i$ to $\mu_j$ if $i\ge 7$ and $j \le 6$. Also, there is no edge from $\{\mu_7,\mu_8,\mu_9\}$ to $\{\mu_{10},\mu_{11},\mu_{12}\}$.} An edge from $\mu$ to $\mu'$ means that $\mu' \unrhd \mu$. 

As can be seen, matchings from $\mu_1$ to $\mu_6$ are covered as they do not dominate any other matching. On the other hand, $\mu_7$ to $\mu_{12}$ are not covered. For example, $\mu_8 \unrhd \mu_7$ and $\mu_7 \unrhd \mu_9$. However, we do not have $\mu_8 \unrhd \mu_9$. Therefore, 

$$ \C = \{\mu_7, \mu_8,\mu_9, \mu_{10},\mu_{11},\mu_{12}\}.$$
\eprf 

\end{spacing}
\end{document}